%**start of header
\newcount\mgnf\newcount\tipi\newcount\tipoformule
\newcount\aux\newcount\piepagina\newcount\xdata
\mgnf=0
\aux=0           %1 produce aux
\tipoformule=1   %0 usa aux; 1 no (usa i simboli dati)
\piepagina=1     %0 =data e #par.#pag; 1=data e #pag; 2=#pag
\xdata=0         %0 data del giorno, 1 data fissa da \Di:
\def\Di{30 aprile 1998}

\ifnum\mgnf=1 \aux=0 \tipoformule =1 \piepagina=1 \xdata=1\fi
\newcount\bibl
%\bibl= ?               % 0= rif [XXX], 1= rif. numerici
\ifnum\mgnf=0\bibl=0\else\bibl=1\fi
\bibl=0

% Per poter cambiare a piacimento il formato dei riferimenti
% bibliografici in <nome>.tex:
%
% 1: citare nella forma esemplificata da \ref{B}{2}{20}}
%    ove XXX e' un simbolo per le iniziali e 2 distingue i lavori con
%    le stesse iniziali,  7 e' il numero SIMBOLICO del riferimento per XXX2.
%    Il numero 7 puo' essere ARBITRARIO e viene automaticamente
%    riaggiustato al momento della compilazione (vedi punto 4)
%
% 2: Se si sceglie \bibl=0 si cita nella forma [XXX2]; se si sceglie
%    \bibl=1 si cita nella forma [numero di ordine di prima citazione].
%
% 3: La bibliografia va scritta nella forma \def{\qqq}{<citazione>}
%    in ordine alfabetico per autore attribuendo un simbolo
%    qualsiasi al testo <citazione} (ad es \def\REFCIT{citazione} e
%    va seguita dal comando  \rif{XXX}{2}{\qqq}{0}
%    scritto all' inizio di una riga altrimenti bianca. L' ultimo {0}
%    e' un parametro inutile ed e' li solo provvisoriamente per memoria
%    (di tentativi in corso).
%
% 4: Per il riaggiustamento automatico (in ordine di citazione)
%    occorre compilare (ottenendo oltre ai soliti la scheda ausiliaria
%    ref.b) e poi si deve eseguire il
%    programma rif <nome> che (usando ref.b) produce fin.tex e <nome>.tex
%    con i riferimenti giusti
%    in \bibl=1 e la si ricompila e stampa. La scheda iniziale <nome>.tex
%    diventa <nome>.old.
\openout8=ref.b
\ifnum\bibl=0
\def\ref#1#2#3{[#1#2]\write8{#1@#2}}
\def\rif#1#2#3#4{\item{[#1#2]} #3}
\fi

\ifnum\bibl=1
\def\ref#1#2#3{[#3]\write8{#1@#2}}
\def\rif#1#2#3#4{}

\fi

\def\9#1{\ifnum\aux=1#1\else\relax\fi}
\ifnum\piepagina=0 \footline={\rlap{\hbox{\copy200}\
$\st[\number\pageno]$}\hss\tenrm \foglio\hss}\fi \ifnum\piepagina=1
\footline={\rlap{\hbox{\copy200}} \hss\tenrm \folio\hss}\fi
\ifnum\piepagina=2\footline{\hss\tenrm\folio\hss}\fi

\ifnum\mgnf=0 \magnification=\magstep0
\hsize=13.5truecm\vsize=22.5truecm \parindent=4.pt\fi
\ifnum\mgnf=1 \magnification=\magstep1
\hsize=16.0truecm\vsize=22.5truecm\baselineskip14pt\vglue5.0truecm
\overfullrule=0pt \parindent=4.pt\fi

\let\a=\alpha\let\b=\beta  \let\d=\delta
\let\e=\varepsilon  \let\h=\eta
\let\th=\vartheta\let\k=\kappa  \let\m=\mu \let\n=\nu
\let\x=\xi \let\p=\pi   \let\t=\tau
 \let\f=\varphi  \let\o=\omega
 \let\G=\Gamma \let\D=\Delta 
   \let\F=\Phi
  
{\count255=\time\divide\count255 by 60 \xdef\oramin{\number\count255}
\multiply\count255 by-60\advance\count255 by\time
\xdef\oramin{\oramin:\ifnum\count255<10 0\fi\the\count255}}
\def\ora{\oramin }

%\Di e' definito all' inizio
\ifnum\xdata=0
\def\data{\number\day/\ifcase\month\or gennaio \or
febbraio \or marzo \or aprile \or maggio \or giugno \or luglio \or
agosto \or settembre \or ottobre \or novembre \or dicembre
\fi/\number\year;\ \ora}
\else
\def\data{\Di}
\fi

\setbox200\hbox{$\scriptscriptstyle \data $}
\newcount\pgn \pgn=1
\def\foglio{\number\numsec:\number\pgn
\global\advance\pgn by 1} \def\foglioa{A\number\numsec:\number\pgn
\global\advance\pgn by 1}
\global\newcount\numsec\global\newcount\numfor \global\newcount\numfig
\gdef\profonditastruttura{\dp\strutbox}
\def\senondefinito#1{\expandafter\ifx\csname#1\endcsname\relax}
\def\SIA #1,#2,#3 {\senondefinito{#1#2} \expandafter\xdef\csname
#1#2\endcsname{#3} \else \write16{???? ma #1,#2 e' gia' stato definito
!!!!} \fi} \def\etichetta(#1){(\veroparagrafo.\veraformula) \SIA
e,#1,(\veroparagrafo.\veraformula) \global\advance\numfor by 1
\9{\write15{\string\FU (#1){\equ(#1)}}} \9{ \write16{ EQ \equ(#1) == #1
}}} \def \FU(#1)#2{\SIA fu,#1,#2 }
\def\etichettaa(#1){(A\veroparagrafo.\veraformula) \SIA
e,#1,(A\veroparagrafo.\veraformula) \global\advance\numfor by 1
\9{\write15{\string\FU (#1){\equ(#1)}}} \9{ \write16{ EQ \equ(#1) == #1
}}} \def\getichetta(#1){Fig.  \verafigura \SIA e,#1,{\verafigura}
\global\advance\numfig by 1 \9{\write15{\string\FU (#1){\equ(#1)}}} \9{
\write16{ Fig.  \equ(#1) ha simbolo #1 }}} \newdimen\gwidth \def\BOZZA{
\def\alato(##1){ {\vtop to \profonditastruttura{\baselineskip
\profonditastruttura\vss
\rlap{\kern-\hsize\kern-1.2truecm{$\scriptstyle##1$}}}}}
\def\galato(##1){ \gwidth=\hsize \divide\gwidth by 2 {\vtop to
\profonditastruttura{\baselineskip \profonditastruttura\vss
\rlap{\kern-\gwidth\kern-1.2truecm{$\scriptstyle##1$}}}}} }
\def\alato(#1){} \def\galato(#1){}
\def\veroparagrafo{\number\numsec}\def\veraformula{\number\numfor}
\def\verafigura{\number\numfig}
\def\geq(#1){\getichetta(#1)\galato(#1)}
\def\Eq(#1){\eqno{\etichetta(#1)\alato(#1)}}
\def\eq(#1){\etichetta(#1)\alato(#1)}
\def\Eqa(#1){\eqno{\etichettaa(#1)\alato(#1)}}
\def\eqa(#1){\etichettaa(#1)\alato(#1)}
\def\eqv(#1){\senondefinito{fu#1}$\clubsuit$#1\write16{No translation
for #1} \else\csname fu#1\endcsname\fi}
\def\equ(#1){\senondefinito{e#1}\eqv(#1)\else\csname e#1\endcsname\fi}
\openin13=#1.aux \ifeof13 \relax \else \input #1.aux \closein13\fi
\openin14=\jobname.aux \ifeof14 \relax \else \input \jobname.aux
\closein14 \fi \9{\openout15=\jobname.aux} \newskip\ttglue

%\font\dodicirm=cmr12\font\dodicibf=cmbx12\font\dodiciit=cmti12
%\font\titolo=cmbx12 scaled \magstep2
%\font\ottorm=cmr8\font\ottoi=cmmi8\font\ottosy=cmsy8
%\font\ottobf=cmbx8\font\ottott=cmtt8\font\ottosl=cmsl8\font\ottoit=cmti8
%\font\sixrm=cmr6\font\sixbf=cmbx6\font\sixi=cmmi6\font\sixsy=cmsy6

\font\titolo=cmbx10 scaled \magstep1
\font\ottorm=cmr8\font\ottoi=cmmi7\font\ottosy=cmsy7
\font\ottobf=cmbx7\font\ottott=cmtt8\font\ottosl=cmsl8\font\ottoit=cmti7
\font\sixrm=cmr6\font\sixbf=cmbx7\font\sixi=cmmi7\font\sixsy=cmsy7

\font\fiverm=cmr5\font\fivesy=cmsy5\font\fivei=cmmi5\font\fivebf=cmbx5
\def\ottopunti{\def\rm{\fam0\ottorm}\textfont0=\ottorm%
\scriptfont0=\sixrm\scriptscriptfont0=\fiverm\textfont1=\ottoi%
\scriptfont1=\sixi\scriptscriptfont1=\fivei\textfont2=\ottosy%
\scriptfont2=\sixsy\scriptscriptfont2=\fivesy\textfont3=\tenex%
\scriptfont3=\tenex\scriptscriptfont3=\tenex\textfont\itfam=\ottoit%
\def\it{\fam\itfam\ottoit}\textfont\slfam=\ottosl%
\def\sl{\fam\slfam\ottosl}\textfont\ttfam=\ottott%
\def\tt{\fam\ttfam\ottott}\textfont\bffam=\ottobf%
\scriptfont\bffam=\sixbf\scriptscriptfont\bffam=\fivebf%
\def\bf{\fam\bffam\ottobf}\tt\ttglue=.5em plus.25em minus.15em%
\setbox\strutbox=\hbox{\vrule height7pt depth2pt width0pt}%
\normalbaselineskip=9pt\let\sc=\sixrm\normalbaselines\rm}
\catcode`@=11
\def\footnote#1{\edef\@sf{\spacefactor\the\spacefactor}#1\@sf
\insert\footins\bgroup\ottopunti\interlinepenalty100\let\par=\endgraf
\leftskip=0pt \rightskip=0pt \splittopskip=10pt plus 1pt minus 1pt
\floatingpenalty=20000
\smallskip\item{#1}\bgroup\strut\aftergroup\@foot\let\next}
\skip\footins=12pt plus 2pt minus 4pt\dimen\footins=30pc\catcode`@=12
\newdimen\xshift \newdimen\xwidth \newdimen\yshift
\def\ins#1#2#3{\vbox to0pt{\kern-#2 \hbox{\kern#1
#3}\vss}\nointerlineskip} \def\eqfig#1#2#3#4#5{ \par\xwidth=#1
\xshift=\hsize \advance\xshift by-\xwidth \divide\xshift by 2
\yshift=#2 \divide\yshift by 2 \line{\hglue\xshift \vbox to #2{\vfil #3
\includegraphics{#4.ps} }\hfill\raise\yshift\hbox{#5}}} \def\8{\write13}

\def\V#1{{\,\underline#1\,}}
\def\T#1{#1\kern-4pt\lower9pt\hbox{$\widetilde{}$}\kern4pt{}}
\let\dpr=\partial \let\io=\infty
\def\fra#1#2{{#1\over#2}}\let\0=\noindent
\def\guida{\leaders\hbox to 1em{\hss.\hss}\hfill}
\def\tende#1{\vtop{\ialign{##\crcr\rightarrowfill\crcr
\noalign{\kern-1pt\nointerlineskip} \hglue3.pt${\scriptstyle
#1}$\hglue3.pt\crcr}}} \def\otto{\
{\kern-1.truept\leftarrow\kern-5.truept\to\kern-1.truept}\ }

\def\pagina{\vfill\eject}\let\ciao=\bye

\def\st{\scriptscriptstyle}
\def\*{\vskip0.3truecm}

\def\lis#1{{\overline #1}}\def\eg{\hbox{\it e.g.\ }}

\def\ie{\hbox{\it i.e.\ }}

\def\fiat{{}}
\def\\{\hfill\break} \def\={{ \; \equiv \; }}
\def\Im{{\rm\,Im\,}} 

\def\annota#1{\footnote{${}^#1$}}
\ifnum\aux=1\BOZZA\else\relax\fi
\ifnum\tipoformule=1\let\Eq=\eqno\def\eq{}\let\Eqa=\eqno\def\eqa{}
\def\equ{{}}\fi
\def\defi{\,{\buildrel def \over =}\,}

\def\1{\ifnum\mgnf=0\pagina\else\relax\fi}
\def\W#1{#1_{\kern-3pt\lower6.6truept\hbox to 1.1truemm
{$\widetilde{}$\hfill}}\kern2pt\,}
\def\Im{{\rm Im}\,}

\def\FINE{
\*
\0{\it Internet:
Authors' preprints downloadable (latest version) at:

\centerline{\tt http://chimera.roma1.infn.it}
\centerline{\tt http://www.math.rutgers.edu/$\sim$giovanni}

\0Mathematical Physics Preprints (mirror) pages.\\
\sl e-mail: giovanni@ipparco.roma1.infn.it
}}

\def\aa{{\V \a}}\def\nn{{\V\n}}\def\AA{{\V A}}
\def\su{{\uparrow}}
\def\oo{{\V \o}}\def\nn{{\V\n}}

\def\ndpr{{\kern1pt\raise 1pt\hbox{$\not$}\kern1pt\dpr\kern1pt}}
\def\Ndpr{{\kern1pt\raise 1pt\hbox{$\not$}\kern0.3pt\dpr\kern1pt}}

\fiat

\def\hdp{{\h^{\fra12}}}\def\hdm{{\h^{-\fra12}}}
\def\su{{\uparrow}}
\font\cs=cmcsc10
\font\msytw=msbm10 scaled\magstep1

\def\RRR{\hbox{\msytw R}}

 \def\ZZZ{\hbox{\msytw Z}}
 
\def\TTT{\hbox{\msytw T}} 

%**end of header
\fiat

\centerline{\titolo Melnikov's approximation dominance. Some examples.}
\*\*

\centerline{\bf G. Gallavotti, G. Gentile, V. Mastropietro}
\*
\centerline{Universit\`a di Roma 1,2,3 }
\centerline{\Di}
\vskip1.truecm

\line{\vtop{
\line{\hskip1.5truecm\vbox{\advance \hsize by -3.1 truecm
\0{\cs Abstract.}
{\it We continue a
previous paper to show that Mel'nikov's first order formula for {\it
part} of the separatrix splitting of a pendulum under fast quasi
periodic forcing holds, in special examples, as an asymptotic formula
in the forcing rapidity.}}
\hfill} }}
\*\*
\0{\bf\S0. Introduction.}
\numsec=0\numfor=1\*

Recently there has been renewed interest in a problem treated in the
paper [G3]. One of the several questions posed in [G3] was to find
upper and lower bounds on the splitting between the stable and
unstable manifolds of the invariant torus with rotation vector
$\oo=\hdm\oo_0$ in the Hamiltonian system (``{\it Thirring model}\/''):
$$\hdm \oo_0\cdot \AA+\fra1{2J} \AA\cdot\AA+
\fra{I^2}{2J_0}+ g^2 J_0 (\cos\f-1)+\m f(\aa,\f) \; , \Eq(0.1)$$
where $(I,\f)\in \RRR\times\TTT,(\AA,\aa)\in
\RRR^{l-1}\times\TTT^{l-1}$ are canonically
conjugated variables, $\oo_0\in \RRR^{l-1}$, and $J,J_0>0$
(respectively, ``rotators' moments of inertia'' and ``pendulum's
moment of inertia''), $g>0$ ($g^2$ is the ``gravity''); $\oo_0,\m$ are
parameters.  Here $J$ could be a scalar or a diagonal
$(l-1)\times(l-1)$ matrix. And setting $\n\=(n,\nn)\in \ZZZ^l$,
$|\n|=|n|+|\nn|=|n|+\sum_{i=1}^{l-1} |\n_i|$:
$$f(\aa,\f)=\sum_{\nn,n;\,|n|\le N_0}
f_{\nn,n}\cos(\nn\cdot\aa+n\f) \; , \Eq(0.2)$$
where $f_\n$ are fixed constants and $N_0>0$ is fixed.\annota1{It
seems clear that the often sought ``first order of perturbation theory''
dominance as $\h\to0$ can only hold if $N_0<\io$ and $f_{\nn,n}$ is
nonzero and ``quite large'' for ``many'' $\nn$'s. Hence
$N_0<\io$ is a reasonable assumption, together with the assumption
that we introduce later on $f_\nn$.}

The paper [G3] did solve completely the upper bound question for $f$ a
trigonometric polynomial: in the sense that it derived generically
optimal bounds for the Fourier transform of the $2$--dimensional
splitting function $\V\D_\su(\aa)$ (a vector valued function 
defined in [G3], p. 366).

Subsequently it was rightly pointed out, [DGJS], that it would be
interesting to just have examples in which some lower bound could be
computed as given by the naive first (non trivial) order perturbation
theory prediction (\ie ``{\it Mel'nikov integral}\/''). For instance
one could study the difference in the free pendulum energy
$h_0=\fra1{2J_0}I^2+J_0 g^2(\cos\f -1)$ evaluated in two
``corresponding points'' of the stable and unstable manifolds of the
invariant torus. Such difference is a very special case of the
splitting functions considered in [G3].

Of course studying the splitting only via the variations of a single
function (``observable'') is very reductive. The stable and unstable
manifolds are $2$--dimensional surfaces, if observed on a
($4$--dimensional) section transverse to the motion and at fixed total
energy, hence it is clear that studying the difference of the 
two manifolds at ``corresponding points'' on the chosen
section requires the simultaneous analysis of $2$ observables.  Thus
in [G3] two independent observables are considered (more are not
necessary: the dimension of the manifolds being $2$).

One also needs to have good control over the Mel'nikov integral, \ie
on the {\it usually explicitly known} first order (in $\e$) expression
of the splitting: not so easy in general. Control is possible in some
rather special cases like the $3$ degrees of freedom system (\ie two
rotators) with $f(\aa,\f)=\sum_\nn f_\nn (\cos
\nn\cdot\aa)\,(\cos\f)$, where $f_\nn= e^{-\k|\nn|}$ or more generally
$F'e^{-\k|\nn|}\le |f_{\nn}|\le F e^{-\k|\nn|}$ for some $F',F,\k>0$,
and $\oo_0=(\o_{01},\o_{02})$ a rotation vector with a golden mean
rotation number $\o_{01}/\o_{02}=\fra12(\sqrt5-1)$.

The problem has been studied, [DGJS], without taking into account [G3]
but without any saving of work because a theory of the upper bound
equivalent to [G3] has to be derived one way or other: and in fact
they had to restrict considerations to the {\it isochronous} cases
($J=+\io$): taking $J<+\io$ {\it does not} change the first order
analysis but changes completely the higher orders treatment.

We show here that the analysis of [G3] yields an
upper bound on the splitting {\it optimal and sufficient} to deduce
the dominance of Mel'nikov's approximation for the part of the
splitting defined by the variation of the function $h_0$ above
defined, if the perturbation is:

$$f(\aa,\f)=\sum_{\nn,\, |\nn|\le N}f_\nn
\cos \nn\cdot\aa\,\cos\f,\qquad N\defi \h^{-1} \; , \Eq(0.3)$$ 
with $f_\nn\ge F e^{-\k|\nn|}$ for some $F,\k>0$ and all $\nn$.\annota2{
This note is made necessary by recent claims that [G3] would not be
sufficient to get an upper bound which would yield the full result of
[DGJS], see [St].}
This differs from the function considered in [DGJS] by a quantity
$O(e^{-\h^{-1}\k})$ which is much smaller than any expected
splitting. The latter is expected to be given by the Mel'nikov
integral, hence to be of order $\ge O(e^{-O(\h^{-\fra12})})$.

Extending the sum over $\nn$ to all $\nn$'s would require extending
the results in [G3] to the case in which the perturbation is analytic,
while one of the ``philosophical assumptions'' of [G3] was that one
should understand first only polynomial cases. However the purpose of
[DGJS] was to provide an example of the Mel'nikov integral dominance
and the \equ(0.3) is as good for the purpose.

Generalizing [G3] to non polynomial analytic cases is easy: but to
limit the length and the {\it simplicity} of the present comment on
[G3] we do not provide the details.  We can indicate that an
extension to analytic cases has been performed in the very
similar problem studied in [BCG] (providing again an example of the
Mel'nikov integral dominance).

The fact that the bounds in [G3] are {\it uniform} in the matrix $J$
(which is summarized in [G3] by calling {\it twistless} the invariant
torus of \equ(0.1) with rotation vector $\oo$) will imply, below, that
the perturbation $f$ in \equ(0.3) also generates a splitting measured
by the variation of the function $h_0$ that is asymptotically exactly
given by Mel'nikov's first order perturbation result for all $J\ge
J_0$. The treatment given below being more general than what would be
needed to just obtain the splitting of $h_0$ also solves ({\it in
part}: see below) the problem posed in [RW]: a paper in which the
basic strategy is adapted from [G3] but which seems to contain
incorrect usage of [G3] leading to errors in intermediate steps and in
the final results, see [GGM3].

The following section {\it assumes full} knowledge of [G3] and we
consider it just as a (conceptually trivial) final comment to it, made
interesting by the idea of [DGJS] of studying lower bounds for single
observables: therefore we call it \S9 (\S10 below contains a few
comments to relate the results to intervened papers), as the paper [G3]
ends rather abruptly (because of exhaustion of the author, and of the
problem) at \S8. We use the notations of [G3] and we freely quote the
formulae in the eight sections of [G3]. The following section has simply
to be regarded as a new section of [G3]. The above is an introduction
and formulae have been labeled by $0$ to avoid confusion with the labels
used in [G3].

The relation between the splitting of the observable $h_0$ and the
$2$--dimensional splitting $\V\D_\su(\aa)$ introduced in [G3] is, on the
Poincar\'e section considered in [G3] (\ie at $\f=\p$, and fixed
energy) and if $J=+\io$, simply $-\oo\cdot\V\D_\su(\aa)$. In the
anisochronous case ($J<+\io$) it is slightly more involved. Unless
explicitly stated, we shall mean that the ``splitting'' is the vector
$\V\D_\su(\aa)$ of [G3], p.366.

\*
\0{\bf\S9. Homoclinic splitting in presence of several fast rotators.}
\numsec=9\numfor=1\*

An interesting consequence of the theory of \S8 is an expression of
the splitting in the $l=3$ case and when the vector $\oo$ has all
components {\it fast}, \ie $\oo=\hdm\oo_0$ and $\h\to0$, while $\oo_0$
is fixed and verifies the Diophantine condition (1.3).

Examining the tree expansion for $\V\D_\su(\a)$ one realizes that
$\V\D_{\su\nn}^h$ is given by a sum of tree values {\it each of which}
is bounded by $\x\=e^{-|\oo\cdot\nn'|g^{-1}(\fra\pi2-d)}$ (where $d$
is prefixed and chosen equal to $\hdp$, as in \S8, to treat the case
$l=2$) with a $\nn'$ in general different from $\nn$ but {\it nonzero},
see p. 379.

Each tree will in general contain several bubbles (see \S6,
p. 376). Here it will be convenient not to lump together the
contributions from the trees with the same {\it free} structure: rather
it will be convenient to leave them separate: thus a tree value will be
described as in \S6 but its fruits will contain their {\it seed} inside
(see \S6, p. 377). We should call them {\it seeded fruits}, for obvious
reasons; below we shall simply call them just ``{\it fruits}'': they can
be either ripe or dry, as defined in \S6, p. 376. Therefore we
redecompose the resummation trees into single trees by specifying the
seed inside each fruit. This is a minor change and it is convenient as
it will show that the bound in (8.1) can be trivially and greatly
improved in the case of perturbations like
\equ(0.3).

The small factor $\x$ arises for some $\nn'$ with $|\nn'|\le Nh$ ($N$
being the {\it cut--off} parameter $N=\h^{-1}$) in two different ways.

\0(1) In trees with free momentum $\nn'$ (see \S6, p.375) and {\it
only} ripe fruits (see \S6, p.376) it arises by bounding carefully
(\ie by complex integration) the integrals relative to the free nodes
(called $v$) variables $\t_v$ and by using the trivial bounds to bound the
fruit values (\ie with no excursion into the complex plane to estimate
them, which of course would not be allowed because of lack of
analyticity of the integrals that express the values of ripe fruits).

\0(2) In trees with at least one dry fruit it arises from a fruit
located inside it, perhaps ``very deeply'' (\ie inside many other
fruits), which only contains ripe fruits. It exists necessarily (note
that no fruit at all is a special case of only ripe fruits) and in this
case the free momentum cannot vanish (see \S6): this was the reason
why the induction described in \S6 worked.

Therefore there is a natural resummation of the splitting series
$\sum_{\nn,h} \e^h e^{i\nn\cdot\aa}\V\D_{\su\nn}^h$: namely we
separate the first order contribution and {\it we collect together all
contributions from trees whose value is bounded by the same small
factor $\x$}. In this way:
$$\V\D_\su(\aa)= \sum_{\nn\neq \V0} e^{i\nn\cdot\aa}
e^{-\fra{\p}{2g}|\oo\cdot\nn|}
\Big( \m \tilde{\V\D}^1_{\su\nn}+\sum_{h=2}^\io
\m^h\tilde{{\V\D}}^h_{\su\nn}(\aa)\Big) \; , \Eq(9.1)$$
where $e^{i\nn\cdot\aa} e^{-\fra{\p}{2g}|\oo\cdot\nn|}
\tilde{\V\D}^1_{\su\nn}$ isolates the first order terms, while
the quantity $e^{i\nn\cdot\aa}$ $e^{-\fra{\p}{2g}|\oo\cdot\nn|}
\tilde{{\V\D}}^h_{\su\nn}(\aa)$ is defined as the sum of the values
$e^{i\nn\cdot\aa} e^{-\fra{\p}{2g}|\oo\cdot\nn|}
\V{{\tilde \D}}^{h,\th}_{\su\nn}(\aa)$ 
of all trees $\th$ of order $h$ containing either
only ripe fruits (or no fruits at all) or a dry fruit which has inside
a seed that is a tree with only ripe fruits (or no fruits at all); see
items (1) and (2) above.

Since a tree can have several fruits which can be ``exponentially
bounded'' there is ambiguity in attributing the terms which contain
several small factors $e^{-{\p\over 2g}|\oo\cdot\nn_j|}$ with
momenta $\nn_1,\nn_2,\ldots$: in the latter cases we just make an
arbitrary choice to attribute such terms to
$\tilde{{\V \D}}^h_{\su\nn}(\a)$ with some $\nn$
among $\nn_1,\nn_2,\ldots$ (\eg $\nn$ is the first among
$\nn_1,\nn_2,\ldots$ in some lexicographic order).

The series involving $\m^h\tilde{{\V \D}}^h_{\su\nn}(\a)$ starts at least at
order two because the trees must be at least of second order, having
already separated out the first order contribution.
Furthermore each tree contains as a factor the product of
the couplings $f_{\nn_v}$ that correspond to its nodes $v$ with order
label $\d_v=1$ (see p. 367); and the set of the $\nn_v$ must be such that
{\it a suitable subset} $W$ of the set of nodes verifies $\sum_{v\in
W}\nn_v=\nn$. This means that we can bound the contribution of each
tree $\th$ with $h$ nodes $v$ of order $\d_v=1$ (and hence total
number of nodes $m\le 2h$) by the bounds in Appendix A1,
in which the part in the constant $c_1$ due to the factors
$f_{\nn_v}$ is explicitly factored out:
$$|\tilde{{\V\D}}^{h,\th}_{\su\nn}(\aa)|\le J_0g D_0 d^{-\b} N^{4h}(B_0
d^{-\b})^{h-1} h!^p\,|f_{\nn_1}|\cdot\cdot\cdot|f_{\nn_h}| \; , \Eq(9.2)$$
where $\b,p$ are the parameters in (8.1),(8.2) with $D_0,B_0$ bounded
in appendix A1, see (A1.2)$\div$(A1.15), by $\lis B$, for some constant
$\lis B$ (which is not the same as $c_3^2M_*^2 N^{-4}$ as in (A1.15)
only because we have factored out the product of the factors
$f_{\nn_v}$ associated with each of the $h$ nodes or order $1$).

Note that in general the label $\nn$ in
$\tilde{{\V\D}}^{h,\th}_{\su\nn}(\aa)$ is not equal to the total
momentum of the tree $\th$ (\ie $\tilde{{\V\D}}^{h}_{\su\nn}(\aa)$
{\it is not} a Fourier component, as it depends explicitly on $\aa$).

It is not difficult to see (by working out a special example, see [GGM3]
for a trivial example; furthermore the proof of the theorem in \S10 of
[CG], although incorrect as a proof of the theorem, nevertheless
provides a much less trivial example) that the above formula and bounds
are optimal, so that a bound like:
$$|\V\D^h_{\su\nn}|< B^{'h} \h^{-\b'}
e^{-\fra{\p}{2g}|\oo\cdot\nn|}\Eq(9.3)$$
for the $\nn$-th Fourier component of $\D_{\su}(\aa)$
would be incorrect.\annota{{6a}}{
The above bound, after summation over $h$, is claimed to hold
in [RW], theorem 2.1, for a quantity that the authors
appeared to think, in private correspondence, to be quite simply
related to the $\V\D_{\su}(\a)$ of [G3] and of this note.}

Looking at \equ(9.1) we can therefore bound the sum over $\nn$ and $h$
of the {\it second} addend by first truncating the series to order $h\le
N$: the remainder is estimated, as in \S8, via the first of
(8.1),\annota{{6b}}{Which descends, for instance, from the KAM--type
theorems in \S5 of [CG], hence it holds for {\it analytic perturbations}
and it is uniform in the perturbation as long as that is analytic in
$|\Im \a_j|<\k$ and $|f_\nn|<F e^{-\k|\nn|}$, if $\k$ is a prefixed
quantity, {\it no matter how small}.}  and will be of the order
$O((\h^{-Q}\m)^N)$ so that if $|\m|<\h^{Q+1}$ it has size not exceeding
$O(\h^{\h^{-1}})$, much smaller than the ge\-ne\-ri\-cal\-ly
expected contribution from the first order terms.

The first $N$ orders can be bounded (as in (8.6) and using $d=\hdp$) by:
$$\eqalign{
& J_0 g  D_0 \h^{-\b/2} N^4
\sum_{h=2}^N \m^h F^h\sum_{\nn_1+\ldots+\nn_h=\nn}
e^{-\k\sum_j|\nn_j|} (B_0 N^4\h^{-\b/2})^{h-1} h!^p\le\cr
& \qquad \le 2 J_0 g D_0 \h^{-\b/2} {c}^{l} N^4 \big(\m F B_0 N^{l+p+4}
l!\h^{-\b/2}\big)^2 e^{-\k|\nn|} \; , \cr} \Eq(9.4)$$
where we have used that $\sum_{\nn_1+\ldots+\nn_h=\nn}
e^{-\k\sum_j|\nn_j|}\le {c}^{l} e^{-\k|\nn|} (lh)^{lh}$ for some
$c$, because the sum has dimension $\le hl$ (in fact $=h(l-1)$) and
we have assumed that $\m$ is so small that $(\m F B_0 N^{l+p+4}
l!\h^{-\b/2})^2$ $<$ $\fra12$. Hence under the condition:
$$|\m|\m_0^{-1}\defi |\m|\,2 D_0 c^l N^4J_0 g \h^{-\b/2}
\left( F B_0 N^{l+p+4} l!\h^{-\b/2} \right)^2<1 \Eq(9.5)$$
the sum in \equ(9.4) is bounded by: $\m e^{-\k |\nn|}$.  Since
$\tilde\V\D^1_{\su\nn}$ is essentially\annota{{6c}}{
Precisely $\V\D^1_{\su\nn}=-i\nn f_\nn \fra{2\pi}{g}\fra{\oo\cdot\nn}
{\sinh \left[ \fra\p2g^{-1}\oo\cdot\nn \right] }=-i\nn f_\nn\F_\nn
e^{-\fra\p2g^{-1}|\oo\cdot\nn|}$
with $1\le |\F_\nn|< (2|\oo\cdot\nn|g^{-1}+1)$.}
$e^{-\k|\nn|}$, we can write
$$ \eqalign{
\V\D_\su(\aa) & =\m\sum_{\nn\ne\V0;|\nn|\le N} e^{i\nn\cdot\aa}
e^{-\fra{\pi}{2g}|\oo\cdot\nn|} \Big(\tilde\V\D^1_{\su\nn}\,+
\m e^{-\k|\nn|}\V \G_\nn(\aa)\Big) \cr 
& +O(\h^{\h^{-1}}) + O(\mu^2 e^{-{\k\over
2}\h^{-1}}) \; , \qquad
|\V \G_\nn(\aa)| <1 \; , \cr} \Eq(9.6)$$
where $O(\h^{\h^{-1}})$ is a bound on the terms with $h> N$ and $O(\mu^2
e^{-{\k\over 2}\h^{-1}})$ is a bound on the terms with $h\le N$ with a
bound proportional to $e^{-\fra{\pi}{2g}|\oo\cdot\nn|}$ with $|\nn|>N$.

By using the expression \equ(9.6) for the splitting vector
we can compute the determinant of the splitting matrix or,
similarly to [DGJS], the variation of the pendulum energy
evaluated on the stable and unstable manifolds, both at $\f=\p$.

In both cases, by conforming to the traditional terminology,
we call Mel'nikov integral the first (non trivial) order
contribution.

If one computes the determinant of the splitting matrix, one has to
differentiate once with repect to $\aa$ the splitting vector
$\V\D_\su(\aa)$. Since the derivative of the terms arising from the
first order contributions is trivial we only need to say that the
derivatives of $\V\G_\nn(\aa)$ can be computed graph by graph and in
each graph they amount to a multiplication by a component of
the total momentum.
Therefore under a slightly stronger requirement on $\m$ (since
$|\nn|<\h^{-1}$ one has to require that $\m$ is smaller than $\h$ times
the $\m_0$ defined above: $|\m|<\m_0\h$) one finds
$$\eqalign{
&\det \dpr_\aa\V\D_\su(\aa) \Big|_{\aa=\V0}=-2\Big(\fra{\p\,\m}{g}\Big)^2
\sum_{|\nn|,|\nn'|\le N}
\fra{\oo\cdot\nn}{\sinh \left[ \fra\p2
\oo\cdot\nn\,g^{-1}\right] }\cdot\fra{\oo\cdot\nn'}{\sinh\left[ \fra\p2
\oo\cdot\nn'\,g^{-1} \right] }\cdot \cr
& \qquad \cdot \Big(f_\nn f_{\nn'}(\nn\wedge\nn')^2+\m 
e^{-\k(|\nn|+|\nn'|)}\G'_{\nn,\nn'}\Big)+O(\mu^2 e^{-{\k\over
2}\h^{-1}}) \; , \cr}\Eq(9.7)$$
with $|\G'_{\nn,\nn'}|<1$, under the stated condition on $|\m|$ (\ie
$|\m|<\h\m_0$).

As to the variation of the pendulum energy, by calling $M(\aa)$ the
Mel'nikov integral (depending on $\aa$, contrary to the {\it
homoclinic determinant} \equ(9.7), which is computed at the homoclinic
point $\f=\p$, $\aa=\V0$), one has
$$ M(\aa) =\m \sum_{\nn\ne\V0;|\nn|\le N} e^{i\nn\cdot\aa}
e^{-\fra{\pi}{2g}|\oo\cdot\nn|}\oo\cdot\V\D^1_{\su\nn} \Eq(9.8) $$
so that, for some constant $C$,
$$ \left|\oo\cdot \V\D_\su(\aa)-M(\aa)\right|
\le C |\m|^2 \sum_{\nn\ne\V0;|\nn|\le N}
e^{-\fra{\pi}{2g}|\oo\cdot\nn|} |\oo| e^{-\k|\nn|} \Eq(9.9) $$
which can be compared with the results in [DGJS], and, unlike [DGJS],
it holds also in the aniso\-chro\-nous case.

{\it Note that so far no use has been made of the assumption that $l=3$
nor that $\oo$ has a rotation number equal to the golden mean nor of the
contents of the paper} [DGJS]. The above \equ(9.8) and \equ(9.7) also seem
well suited for extensions to $l>3$ (or to more general
rotation numbers) of the Mel'nikov integral dominance.
\*

\0{\it Remark.} Assuming $f_\nn= e^{-\k|\nn|}$ for $|\nn|<N$, by an
argument of [DGJS] (exposed in \S6 of [DGJS], see also \S2) we see
that, if $l=3$ and $\oo_0$ has rotation number equal to the golden mean
(${\o_{01}}/{\o_{02}}=\fra12(\sqrt5-1)$), in
\equ(9.8) the Mel'nikov integral dominates the asymptotics as
$\h\to0$ if $\m$ is small as indicated. In fact, apart for some
``special'' small intervals of values of $\h$ and of $\a$, {\it only two
addends matter} in \equ(9.8) (see [DGJS]): they correspond to $\nn$
equal to $\pm\nn_k=\pm(f_{k+1},-f_{k})$ where $f_j$ is Fibonacci's
sequence and $k=k(\h)$ is the value that minimizes
$$ \k|\nn_j|+\fra{\p}{2}\hdm|\oo_0\cdot\nn_j|\Eq(9.10)$$
(of course $|\nu_k|\le N$): the corresponding miminum has the form
$c(\h)\h^{-\fra14}$, with $c(\h)>0$ an explicitly computable quantity
bounded above and below by positive constants, uniformly in $\h\to0$.
The special values of $\h$ are those for which the minimum is taken over
{\it two} values of $k$, or of $\h$'s close enough to such values; the
special values of $\aa$ are those for which $|\sin\aa\cdot\nn_k|<b$,
where $b>0$ is an arbitarily prefixed quantity (if $b$ is too small then
the size of $\h$ for which dominance occurs becomes correspondingly
smaller).  Excluding the just mentioned values of $\aa$ and $\h$ the
asymptotics of the Mel'nikov integral $M(\aa)$ is proportional to
$\m\hdm e^{-c(\h)\h^{-\fra14}}\sin\nn_k\cdot\aa$.  Moreover to the
r.h.s. of
\equ(9.9) the same considerations trivially apply,
\ie only two (or four) addends matter so that it is bounded
by a constant times $\m^2 \hdm e^{-c(\h)\h^{-\fra14}}$ and the 
first order dominance for $\oo\cdot \V\D_\su(\aa)$ follows, 
for all $\aa$ except the special ones. So [G3] and 
the first order analysis (in [DGJS]\annota{{6d}}{The rest
of the paper [DGJS] provides an interesting partial alternative to the
work in [G3]; the partiality being mainly due to the isochrony
assumption, essential in the theory of [DGJS].}) prove 
the first order dominance not only in the isochronous case,
as done in [DGJS], but also in the anisochronous case.

\*

But of course the really interesting quantity is \equ(9.7): \ie the
actual splitting at $\aa=\V0$. For this quantity we {\it cannot}
conclude dominance of the first order term, unless some ({\it likely},
see [GGM1], but yet undiscovered) cancellations take place: i{\it t is
manifest from \equ(9.7) that the would be leading term
($\nn=\nn'=\nn_k$) vanishes}.  Hence no conlusion can be drawn, in the
class of models considered here, about the splitting at the homoclinic
point (besides the one already derived in \S8 and saying that it is
smaller than any power of $\h$). Unless one considers $\h\to0$ along a
special sequence in correspondence of which there are {\it two pairs} of
successive Fibonacci's vectors with the same, or almost the same,
$\k|\nn_k|+{\p\over2}|\oo\cdot\nn_k|$: in this case, obviously, we have
dominance of the Mel'nikov term because no two Fibonacci's vectors are
parallel, see \equ(9.7).  Note that the expressions ``almost the same''
and ``close enough'', used above for the variable $\h$, simply mean that
the two Fibonacci's vectors such that the quantity
$\b_k=\k|\nn_k|+{\p\over2}|\oo\cdot\nn_k|$ is closest to its minimum
over $k$ for two values of $k$, say $k,k+1$, and $\b_k-\b_{k+1}$ is a
quantity that is uniformly bounded as $\h\to0$.
\*

\0{\bf\S10. Comments.}
\*
\0(1) The above theory is different from that of [DGJS] in
a radical way; in fact it works in the isochronous case ($J=+\io$) as
well as in the anisochronous one. {\it This is a new result} (see
however the acknowledgements below). And the last remark shows that
the problem of the homoclinic splitting asymptotics with $2$ fast
forcing frequences is still open if it is intended as defined by
\equ(9.7) (as one should in view of the possible applications to
heteroclinic chains and Arnol'd diffusion). This is so in spite of
apparent claims to the contrary, [St].

\*
\0(2) In {\it three} time scales problems (\ie
when $\oo=(\hdm\o_{01},\hdp\o_{02})$ and $g=O(1)$) the above arguments {\it
do not lead to any conclusion} because $\oo\cdot\nn$ is not always
proportional to $\hdm$ but it can {\it even} be proportional to $\hdp$
also for small $|\nn|$: so that the exponentially small factor is not
always present. 
This means that {\it even the individual components of
the splitting $\V\D_\su(\aa)$ in general are not only not dominated by
the Mel'nikov integral but have size of the order of a power of $\h$}!

Regarding the intersection matrix, {\it if} \equ(9.3) {\it could be
assumed valid three matrix elements would be exponentially small} and
also the determinant would be (exponentially small and) 
dominated by the Mel'nikov integral. But, as we said,
\equ(9.3) is false.  In fact the intersection
matrix has, in this case, {\it all matrix elements that have polynomial
size} as $\h\to0$, starting with second order: this part of the theorem
in \S10 of [CG] is, of course, valid.\annota{{6e}}{The theorem was
``proved'' by showing that three matrix elements had polynomial size
while the fourth was ``exponentially small'': the last claim is
incorrect because of a trivial computational error which, if corrected,
yields that even the fourth matrix elements is of polynomial size.} And
it is perhaps natural to think that {\it also} the splitting \equ(9.7)
has size of order of a power of $\h$.  This is however {\it an error},
as the determinant is indeed {\it exponentially small} in $\h$, {\it
although the matrix elements have size of the order of a power of $\h$},
due to the presence of cancellations in the determinant.  This was
proved in [GGM1], where such cancellations are exploited.

The above error was in fact suffered in [CG]: a {\it trivial
computational error} in one of the about twenty terms contributing to
the third order crept in and led the authors of [CG] to believe they
had checked that {\it even the determinant of the splitting matrix}
would be polynomially bounded away from zero as $\h\to0$ (\ie bounded
as the individual matrix elements are) under the convergence condition
that $\m$ is itself a (large) power of $\h$. In any case these
considerations show that there is apparently no hope, without
exploiting cancellations in the determinant, to obtain a exponentially
small bound (the matrix elements have polynomial size); the derivation
of the exponentially small bounds for this case from \equ(9.3), to
which many authors seems apparently to believe, is false because \equ(9.3)
is false, in general and on the section considered here.
\*
\0(3) But the paper [CG] {\it had, {\it nevertheless}, laid the foundations
of the theory developed in} [G3]: a very simple and systematic theory,
because of its field theoretic viewpoint, and very flexible thanks to
the possibility of rapid (graphical) comparison of arbitrarily high
order terms of perturbation expansions. Enough to cope with the error
and to correct it, as shown in [GGM1] (an achievement scarsely
appreciated in [St]). In fact the papers [G3], [GGM1] suggest which
could be the general solution to the problem of the splitting \equ(9.7)
{\it even when the interaction is a trigonometric polynomial}: see the
conjecture in [G4]; a ``few'' cancellations remain to be checked to
prove (or to disprove) the conjecture.

In fact the reason the error was not spotted by the authors of [CG]
(it was pointed out by Gelfreich) was due to the {\it apparent}
matching of the result with the intuitive idea that the splitting in
three time scale problems would be ``large''. The splitting,
defined as the smallest eigenvalue of the intersection matrix, has the
dimension of an action (being the derivative with respect to an angle
of the difference in the actions for the stable and for the unstable
manifolds) {\it hence it has to be measured with respect to an
action}. In [CG] the relevant action with which to compare the
splitting was the size of the gaps, in action space, between the
average actions of the invariant tori surviving the perturbation.

The latter gap was estimated in [CG] by a power, as large as wished,
of $\h$ times $\max\{J_0g,Jg\}$: hence the splitting was larger than
any power of $\h^{-1}$ {\it with respect to the gaps}.

The proof was wrong and it was corrected in [GGM1]. And in [GGM2] it
has been shown (to the skeptics, see [St])\annota{{6f}}{Extending a
property and a method that was noted long ago by Neishtadt, [Ne].}
that the gaps have {\it in this case} size of almost $O(e^{O(-\h^{-1})})$
while the splitting is dominated by the Mel'nikov integral {\it for
generic perturbations}, even if of polynomial type: the Mel'nikov
integral being of order $O(e^{-\hdm\fra{\p\o}{2g}})$ if the frequency
vector is $\oo=(\hdp\o_1,\hdm\o)$. So the splitting is still very
large.\annota{{6g}}{And {\it even larger than value claimed} in [CG]:
the ratio between the splitting and the gap being now
$O(e^{O(\h^{-1})})$, while in [CG] it was claimed to be ``only''
$O(\h^{-a})$ for arbitrary $a$.}

\*

\0(4) Recently we learnt of Eliasson's remark that
the homoclinic splitting $\V Q(\aa)$ is a {\it gradient} $\V
Q(\aa)=\V\dpr_\aa \F(\aa)$ of some {\it potential} $\F$: this is due
to the Lagrangian nature of the stable and unstable manifolds. Once this
is known one recognizes that the tree expansion in sections 1$\div$8 of
[G3] provides a power series expansion of the potential $\F$, see [G4].

\*

\0(5) One can apply the above theory to the case in which the
dimension of the angles $\aa$ is $1$: this is the periodic forcing
case discussed in \S8. In this case there is no distinction between
the \equ(9.7) and \equ(9.8) and one finds an asymptotic expansion for
such splittings (see \S8). This case is closely related to a
corresponding problem in the theory of the standard map: the basic
approach is due to Lazutkin and recently there has been renewed
interest in it. The paper [Ge], which is nice and accurate, also
fills some (apparently) missing points in the original work. The
approach in [Ge] is also based on formal power series expansions for
the separatrix; but unlike [G3] the control of the remainders is not
solely made via the analysis of formal series, thus providing an
interesting alternative, although (of course) the amount of work is
comparable.

The drawback, at the moment, is the need of an ``integrable normal
form'' near the invariant tori (circles in this case). Therefore,
although adapting the method of [Ge] to cover the cases of Hamiltonian
systems appears straightforward in the case of isochronous systems
(which trivially admit an integrable normal form), a possible extension
to the anisochronous case seems to require substantial extra work, at
least if one proceeds along the lines of [Ge]. The method of [G3] has
not (yet) been applied to study the standard map case.

Likewise the extension to higher dimension seems to require further
extra work (a beginning of the theory is in [DGJS]); and so does the
extension to three time scales problems. 

Once the above extensions will exist they will be be very interesting,
and, solving in a substantially different way the problems solved in
the latter papers, they will permit a (better) comparison with the
methods in [G3] and [GGM1].
\*

\0(6) Finally the reason why one may hope to get large {\it
absolute}\annota{{6h}}{{\it I.e.} large compared to the natural unit of
action $J_0g$ and not compared to the size of the gaps.} splitting in
three degrees of freedom systems is {\it not} because some $\oo\cdot\nn$
(in the small factors $e^{-|\oo\cdot\nn|\fra\pi{2g}}$) becomes small
when $\nn$ becomes a good rational approximation of $\oo^\perp$, as
sometimes hinted in the literature, see [St]. This is shown by the three
time scales problem in (2) above in which $\oo\cdot\nn$ can be very
small even for small $\nn$ (\ie if the fast component of $\nn$ vanishes)
and {\it nevertheless} the splitting does not have a polynomial absolute
size, due to remarkable cancellations (see [GGM1]).  One has rather to
think that the splitting can become large because the perturbation
$f(\aa,\f)$ has a small analyticity band in complex $\f$ plane: in fact
the higher the degree $N_0$ in $\f$ the smaller it seems that one has to
choose $\m$ with respect to $\h$ if one wants to guarantee Mel'nikov's
approximation to dominate.
\*

\0{\bf Acknowledgements. \it The content of this paper was (and is) 
considered by us a trivial remark on the work [G3]: as one can see in
the story [St] this view is violently not shared by other
specialists. Therefore we wrote it up to clarify our methods and
techniques, which in spite of apparently preconceived claims to the
contrary (see [St]), are proving more and more well suited for the
problems under analysis. We are greatly indebted to G. Benfatto,
G. Benettin and A. Carati for clarifying discussions and for support.}
%\pagina

\*
\0{\bf References.}
\*
\item{[BCG] } Benettin, G., Carati, A., Gallavotti, G.: {\it A rigorous
implementation of the Jeans--Landau--Teller approximation for
adiabatic invariants}, Nonlinearity {\bf 10}, 479--507 (1997).

\item{[CG] } Chierchia, L., Gallavotti, G.:
{\it Drift and diffusion in phase space},
Annales de l'In\-sti\-tut Henri Poincar\'e B {\bf 60}, 1--144 (1994).

\item{[DGJS] } Delshams, S., Gelfreich, V.G., Jorba, A., Seara, T.M.:
{\it Exponentially small splitting of separatrices under fast
quasiperiodic forcing}, Communications in Mathematical Physics
{\bf189}, 35--72 (1997).

\item{[Ge] } Gelfreich, V.: {\it A proof of the exponentially small
transversality of the separatrices of the standard map}, mp$\_$arc@
math. utexas. edu, \#98--270.

\item{[G3] } Gallavotti, G.:
{\it Twistless KAM tori, quasi flat homoclinic intersections, and
other cancellations in the perturbation series of certain completely
integrable Hamiltonian systems. A review}, Reviews on Mathematical
Physics {\bf6}, 343--411 (1994).

\item{[G4] } Gallavotti, G.: {\it Reminiscences on science at I.H.E.S. 
A problem on homoclinic theory and a brief review}, preprint, 1998,
http://ipparco.roma1.infn.it.

\item{[GGM1] } Gallavotti, G. , Gentile, G., Mastropietro, V.:
{\it Separatrix splitting for systems with three degrees of freedom},
Preprint, in mp$\_$arc@math.utexas.edu, \#97-472 with the title {\it
Pendulum: separatrix splitting}.

\item{[GGM2] } Gallavotti, G., Gentile, G., Mastropietro,V.:
{\it Hamilton-Jacobi equation, heteroclinic chains and Arnol'd
diffusion in three time scales systems}, archived in
chao-dyn@xyz. lanl. gov \#9801004.

\item{[GGM3] } Gallavotti, G., Gentile, G., Mastropietro,V.:
{\it A comment on the Physica D paper by Rudnev and Wiggins},
mp$\_$arc@math.utexas.edu, \#98--245. See also {\it Homoclinic
splitting, II. A possible counterexample to a claim by Rudnev and
Wiggins on Physica D}, chao-dyn 9804017.

\item{[Ne] } Neishtadt, A.I.: {\it The separation of motions in
systems with rapidly rotating phase}, Journal of Applied Mathematics
and Mechanics {\bf48},(2), 133--139 (1984).

\item{[RW] } Rudnev, M.,  Wiggins, S.: {\it Existence of exponentially
small sepratrix splittings and homoclinic connections between whiskered
tori in weakly hyperbolic near integrable Hamiltonian systems},
Physica D {\bf114}, 3--80 (1998).

\item{[St]} Collection of the comments received by us from colleagues
and from referees about the paper [GGM1], and our replies: in
http://ipparco.roma1.infn.it.

\def\FINE{
\0{\it Internet:
Author's preprints downloadable (latest version) at:

\0{\tt http://ipparco.roma1.infn.it}

%\0 mirror: {\tt 
%http://www.math.rutgers.edu/$\sim$giovanni}

\0\sl{}e-mail: Giovanni.Gallavotti@roma1.infn.it,
Guido.Gentile@roma1.infn.it,\\ Vieri.Mastropietro@roma1.infn.it }
\annota{{\bf{}^*}}{\0G.Ga.: Dipartimento di Fisica, Universit\`a 
di Roma 1, P.le Aldo Moro 2,
00185, Italy\\
\0G.Ge.: Dipartimento di Matematica, Universit\`a di Roma 3, Largo
S. Leonardo Murialdo 1, 00146, Roma, Italy\\
\0V.Ma.: Dipartimento di Matematica, Universit\`a di Roma 2, V.le della
Ricerca Scientifica, 00133, Roma, Italy
}}

\*
\FINE
\*
\0Paper archived  in
mp$\_$arc@math.utexas.edu 98-???

\0and chao-dyn@xyz.lanl.gov 98?????
\*

\ciao